**CIVILTA' DELLE MACCHINE**

# EINSTEIN E IL RINNOVAMENTO DELLE SCIENZE:

# 1978 N. 3-6 - Maggio - Dicembre

Erasmo Recami

<<

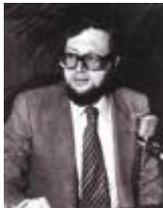

*Erasmo Recami, nato a Milano nel 1939, si e' laureato in fisica a Milano nel 1964 ed ha preso la libera docenza in fisica teorica (Roma, 1971). Attualmente e' professore di fisica superiore all'Universita' di Catania, e collaboratore (col grado R-3) dell'Istituto Nazionale di Fisica Nucleare. E' autore di quasi un centinaio di pubblicazioni scientifiche nei settori: della relativita', della fisica delle particelle, e della meccanica. Ha inoltre scritto numerosi articoli semidivulgativi o didattici o storici, e contribuito a vari volumi di editori internazionali specializzati, a enciclopedie, ecc. Ha svolto ricerca scientifica, oltre che in Italia, anche negli USA, in URSS, in Inghilterra, Danimarca, Polonia, Brasile, ecc.*

Il genere *Homo* è apparso sulla terra molti milioni di anni fa, come suggeriscono le recenti scoperte del 1975 e 1976 effettuate ad Hadar (1), in Etiopia, e a Leatolil (2), in Tanzania. E da quando l'uomo esiste pare che abbia sempre manifestato un comportamento in cui sono riscontrabili componenti scientifiche (all'inizio tecnologiche), artistiche, magico-religiose, e così via. Un esempio ce lo forniscono certe amigdale in pietra ascendenti a quasi un milione di anni fa, in cui la stessa ricerca della perfezione simmetrica non ha, né può avere, fini semplicemente utilitaristici e funzionali.
In ogni caso l'uomo ha sempre agito sulla natura, e osservato razionalmente la realtà, mediante gli strumenti sensoriali e logici ereditati da oltre tre miliardi di anni di evoluzione biologica - evoluzione avvenuta in presenza di una continua interazione tra il vivente e il mondo esterno -. Il maggiore risultato di questo continuo sforzo non solo di operare sulla realtà ma anche di comprendere i meccanismi di tale azione (o, meglio, interazione), nonché di razionalizzare e socializzare il mondo delle proprie sensazioni per arricchire la comunicazione con i propri simili, è stato la creazione del linguaggio: creazione in cui è facile riconoscere il contributo di tutte le facoltà umane, e tra i primi il contributo dell'atteggiamento scientifico (si pensi al carattere «inter-soggettivo» del linguaggio). In un certo senso la scienza moderna non fa altro che ampliane (con telescopi o microscopi) e approfondire (con misurazioni precise) il campo della nostra esperienza, per costruire quindi un supplemento di linguaggio atto a descrivere (e «capire»«) i nuovi mondi di esperienza. Da questo punto di vista fattività scientifica dell'uomo nei millenni presenta forti caratteri di continuità. Essa, inoltre, è in una certa misura «spontanea» e universale: già da poppanti (e ciò è interessante quando si ricordi che l'ontogenesi ripete la filogenesi) di fronte al fenomeno della gravità cominciamo tutti a fare prove e riprove - per noi utili e necessarie - sul modo di agire di questa forza...
Successivamente alcuni rallentano l'attività scientifica; altri, invece, non si accontentano e la continuano, per di più interiorizzandola e mutandola da istintiva in razionale ed esplicita.

La scienza non è, poi, nettamente separabile dalle altre attività «umanistiche». «La scienza, come l'arte, è un'opera umana in pieno sviluppo, ed a carattere dinamico», per dirla con le parole del Premio Nobel Kastler. La scienza è una creazione dello spirrtó umano;, direi anzi una «libera creazione», nello stesso senso in cui sono libere ad esempio le arti o le matematiche. La scienza, infatti, è vincolata (oltre che dalla stessa struttura della nostra mente, del nostro cervello, quale si è sviluppata - come detto - durante l'evoluzione) dal confronto con la realtà esterna; ma anche le arti, e le matematiche, ad esempio, sono vincolate dal confronto con realtà e strutture che grosso modo si possono dire «interiori» e che non hanno meno «oggettività» di quelle esterne. Anzi, tanto le scienze quanto le arti e le matematiche devono cercarsi valori inter-soggettivi per essere riconosciute dalla comunità umana (accomuno sempre arti e matematiche perché non vedo differenze sostanziali tra musica o architettura da una parte, e matematica dall'altra parte). Einstein stesso dichiarò nel 1933: «I postulati e i concetti fondamentali sui quali si basa la fisica teorica sono libere invenzioni dell'intelletto umano..., e costituiscono la parte essenziale di una teoria, la parte che la logica non può toccare».
La scienza, naturalmente, viene sviluppandosi con caratteristiche proprie, soprattutto di metodo, ma essa richiede doti di creatività, intuizione, fantasia, senso estetico non meno che di intelligenza, logica, coerenza, spirito critico, spirito di osservazione quantitativa, precisione, onestà, costanza, abilità «tecnica» in senso sperimentale o teorico-matematico, e così via. Il moderno pensiero scientifico (purtroppo rimasto misconosciuto, e non ancora diffusosi in un «umanesimo scientifico») da molti decenni ha al suo attivo contenuti culturali, formativi, filosofici, logici, che sono forse il prodotto più caratteristico e l'eredità del nostro secolo. Ma, come si è cercato di dire, la scienza al pari degli altri prodotti dell'uomo si è sempre sviluppata con continuità durante l'evoluzione umana; anche se un caratteristico errore prospettico ci fa sembrare tanto più lento il progresso quanto più esso è antico(3). Semmai si può dire che, soprattutto dalla fine del diciassettesimo secolo, la scienza ha assunto con evidenza il proprio carattere di essere una costruzione fondata sulla cooperazione di molti: la collaborazione di molti studiosi, con la loro naturale varietà di interessi e talenti, produce quella messe di osservazioni, calcoli, idee che permette i progressi successivi e che costituisce l'indispensabile base di partenza per ogni importante conquista (compresa quella del genio). Ciò ara stato compreso molto bene da Seneea, il quale - dopo avere discusso delle comete, su cui aveva delle idee molto moderne - concluse non essere ancora giunto il momento per un giudizio definitivo, con le nobili parole: «Accontentiamoci di ciò che abbiamo finora scoperto. Quelli che verranno dopo di noi aggiungeranno il loro contributo alla verità». Anche Einstein non considerò le proprie equazioni della relatività generale come l'ultima parola, chiamandole anzi «effimere» (4), e in una sua lettera privata scrisse (5): « Tu immagini che io guardi indietro alla mia vita con calma soddisfazione. Ma da vicino le cose appaiono diversamente. Non c'è un singolo concetto della cui incrollabilità io sia convinto, e non so se in generale io sono sulla strada giusta .. ».

Anche i rinnovamenti scientifici, quindi, che sono il risultato di crisi di crescenza, possono dirsi avvenire nella continuità. Avvicinandoci alla gigantesca figura di Einstein dobbiamo chiederci non tanto quali rivoluzioni metodologiche egli abbia introdotto, quanto piuttosto come la sua opera e il suo esempio (apportatore di una grande ventata di freschezza) ci abbiano ricondotto alle vere sorgenti e alle eterne caratteristiche della grande scienza. Il loro ricordo spesso sbiadisce nei lunghi anni che la scienza suole spendere per sistemare, completare, applicase le teorie esistenti e per raccogliere gli elementi che porteranno alle teorie nuove.
Ciò detto, bisogna però osservare che la stessa consapevolezza che la scienza ha di sé e del proprio metodo si evolve ed arricchisce, cosicché l'esempio che Einstein ci ha porto attraverso la sua opera e vita scientifica è stato fondamentale per indirizzare - nei fini e nei mezzi - l'intera fisica del nostro secolo, e probabilmente del futuro.
Einstein ci ha ricordato che fare scienza vuol dire non certo redigere un catalogo di fatti (anche se questo può essere un necessario passo preliminare), bensì «capirli» scoprendo ciò che resta costante nel loro manifestarsi e divenire, e facendo quindi discendere le leggi di conservazione così scoperte da un minimo di principi logici e di ipotesi di simmetria. Einstein, anzi, un giorno osservò che non poteva capire «come mai qualcuno potesse sapere così tante cose, e capire così poco» (6) e sempre sottolineò il pelicolo di conoscere troppi fatti e di perdersivi in mezzo. Invece la scienza tende a collocare i fenomeni in strutture ordinate, così da poterli descrivere in una maniera infinitamenta più elegante, logica, sintetica e compatta che non «per elenco». A tale scopo non sovvengono molto le deboli forze della tanto teorizzata induzione. Einstein dichiarava nelle sue Note Autobiografiche (7): «Una teoria può essere verificata dall'esperienza, ma non esiste alcun modo per risalire dall'esperienza alla costruzione di una teoria. Equazioni di tale complessità, come sono le equazioni del campo gravitazionale, possono essere trovate solo attraverso la scoperta di una condizione matematica logicamente semplice, che determini completamente o quasi completamente le equazioni. Una volta in possesso di condizioni formali abbastanza stringenti, non c'è bisogno di una grande conoscenza dei fatti per costruire una teoria...».

Aggiungiamo che Einstein ricorse spesso addirittura ad esperimenti puramente «concettuali» (*gedankenexperimente*), mostrandone la potenza euristica.

Einstein ci ha ricordato - col suo stesso esempio - che la scienza per rinnovarsi ha bisogno della fresca novità delle idee dei giovani. Noi siamo abituati a immaginarci Einstein col maglione e i capelli bianchi: ma gli anni «più incredibili» della sua produzione scientifica sono il 1904 e 1905, quando egli aveva venticinque anni. Paolo Straneo (già professore emerito all'Università di Genova, nonché suo amico) mi ha testimoniato che Einstein già a diciassette anni gli poté esporre chiaramente, passeggiando sulla riva del Lago di Zurigo, quella problematica la cui soluzione lo avrebbe portato otto anni dopo a formulare la Relatività Speciale. Nel 1921 Einstein stesso scrisse a un amico: «La scoperta in grande stile è fatta per i giovani.., e di conseguenza costituisce per me una cosa del passato». La scuola tende à volte a squadrare troppo la mente dei giovani in una forma preconfezionata. Einstein ebbe la fortuna di possedere fortissimo l'istinto di proteggere la originalità della sua mente, prima ancora che del suo pensiero. Si lamentò del ginnasio Luitpold, di Monaco, «a causa del metodo di insegnamento noioso e meccanizzato. Tenuto conto della mia scarsa memoria per le parole, esso mi causava gravi difficoltà e mi sembrava insensato tentare di sormontarle. Preferivo pertanto subire punizioni di ogni genere piuttosto che imparare a farfugliare a memoria». Einstein però studiava molto, anche se per conto suo: «Il mio maggior difetto consisteva nella scarsa memoria, soprattutto per parole e testi... Soltanto in matematica e in fisica ero - grazie ai miei studi personali - molto più avanti del programma scolastico, e altrettanto può dirsi per quanto concerne la filosofia, sispetto al programma della Scuola» (8) (a proposito di filosofia, Einstein a tredici anni aveva già letto - e capito - le opere di Kant). Dirà ancora, anche se con riferimento alla musica: «Io ritengo, tutto, sommato, che l'amore sia un maestro più efficace del senso del dovere». Se mi si concede una divagazione, Einstein fin da bambino - così come rifuggiva istintivamente dalle coercizioni - non apprezzava la vista e i suoni delle sfilate militari del suo Paese; nel 1933 dichiarò: «La mentalità militare troppo accentuata dello Stato tedesco mi era estranea fin dalla fanciullezza. Quando mio padre si trasferì in Italia fece i passi necessari - dietro mia richiesta - per liberarmi dalla cittadinanza tedesca, in quanto volevo diventare cittadino svizzero» (8). In seconda ginnasio, anzi, troncò gli studi scolastici per un anno raggiungendo la famiglia a Milano, e vagabondando ad esempio - sembra a piedi - fino a Genova, ove aveva dei parenti (che lo aiuteranno finanziariamente durante gli anni dell'università). Riprese poi gli studi, ma in Svizzera (otterrà la cittadinanza elvetica nel 1901, conservandola per tutta la vita). Dei quattro anni di studio all'ottimo Politecnico di Zurigo ha lasciato scritto (7): «Per superare gli esami, volenti o nolenti, bisognava imbottirsi la mente di tutte queste nozioni... Una simile coercizione ebbe su di me un effetto così scoraggiante che, dopo aver superato l'esame finale, per un anno intero mi ripugnò prendere in considerazione qualsiasi problema scientifico. E' un vero miracolo che i metodi moderni di istruzione non abbiano ancora completamente soffocato la sacra curiosità della ricerca: perché' questa delicata piantina, oltre che di stimolo, ha soprattutto bisogno di libértà... E' un gravissimo errore pensare che la gioia di vedere e di cercare possa essere suscitata per mezzo della coercizione e del senso del dovere». Einstein si difese anche dalle mode e dai pregiudizi della stessa ricerca scientifica accademica («Nel campo di coloro che cercano la verità - scrisse (8) nel 1953 - non esiste alcuna autorità umana. Chiunque tenti di fare il "magistrato", viene travolto dalle risate degli dei»). Senza il suo sospetto per le autorità costituite forse non avrebbe potuto mantenere intatta quella sua formidabile indipendenza di giudizio che gli diede il coraggio (perché di vero e proprio coraggio eroico si tratta) di contraddire alcune delle più diffuse scuole scientifiche del suo tempo. In questo fu aiutato - se così si può dire - dalla sorte apparentemente maligna, che lo tenne lontano dalle università' a partire dalla sua laurea (1900) fino al 1909, cioè per ben nove anni. Al termine dei corsi universitari, all'età di ventun anni, chiese infatti un qualsiasi posto all'Università, ma venne respinto (8): «Stando a quanto mi riferisce la gente, non sono nelle buone grazie di nessuno dei miei ex-professori», scrisse nel 1901. Mentre, per mantenersi, dava lezioni private a Zurigo, dopo di avere pubblicato il primo articolo scientifico (sulla capillarità) si rivolse per lettera all'illustre Prof. W. Ostwald: «Poiché sono stato ispirato dal Suo libro sulla chimica generale a scrivere l'allegato articolo, mi permetto di mandarGliene una copia. Con l'occasione, mi azzardo inoltre a domandarLe se per caso non potrebbe essere utile un fisicomatematico che ha familiarità con le misurazioni assolute. Mi prendo la libertà di farLe una simile richiesta solo perché sono privo di mezzi e solamente un incarico del genere mi consentirebbe di perfezionarmi ulteriormente negli studi» (8). Non ottenne neppure risposta, al punto che lo stesso padre di Albert Einstein, all'insaputa del figlio, osò rivolgersi a1 famoso Accademico scrivendo tra l'altro: «Mio figlio è profondamente amareggiato dall'attuale inattività e ogni giorno di più si radica in lui l'idea di avere fallito nella propria carriera e di non potere più ritrovare la via giusta. Per giunta, lo deprime il pensiero di esserci di peso... Se potesse esserLe possibile trovargli un posto di assistente, o subito o nel prossimo autunno, la mia gratitudine non avrebbe limiti» (8). Un anno dopo, nel 1902, come sappiamo, ottenne (per interessamento del compagno di studi M. Grossmann) il suo impiego all'Ufficio Brevetti di Berna. Questo ufficio gli offrì la protezione necessaria per portare a compimento i lavori fondamentali apparsi poi nel 1905. Einstein lo definì in una lettera privata «quel chiostro secolare ove si sono dischiuse le mie più belle idee». La originalità delle idee di Einstein, dunque, non fu turbata dalle regole dell'ambiente accademico; eppure non gli mancò lo scambio delle idee perché poté contare sulla colta amicizia, oltre che di Grossmann, anche di un giovane matematico (K. Habicht), di uno studente di filosofia (M. Solovine), e del giovane ingegnere italiano Michele Besso (l'unico che Einstein ringrazi alla fine del suo articolo del 1905 sulla Relatività Speciale). La massima fortuna di Einstein, comunque, fu di potere scrivere in tedesco (allora la lingua della fisica teorica) ed inviare i suoi manoscritti ad una delle maggiori riviste di fisica del tempo, e più ancora di avere trovato sempre immediata aocoglienza - pur non appartenendo ad alcuna università - presso la direzione degli *Annalen der Physik*, capeggiata dal grande Planck; ciò torna a grande onore di Planck e del suo comitato redazionale.

Einstein aveva studiato per conto suo praticamente tutto quanto di importante c'era in fisica al proprio tempo - guidato dal suo intuito quasi magico; ma un'idea di quanto il giovane Einstein fosse al di fuori della consueta circolazione d'idee universitaria ci è forse data da un ben noto fatto relativo alle sue ricerche giovanili, ricordato da lui stesso (7): «Non conoscendo le precedenti ricerche di Boltzmann e Gibbs che avevano già esaurito l'argomento, sviluppai la meccanica statistica e la teoria cinetico-molecolare della termodinamica che si basava su di essa. Il mio scopo precipuo era di trovare fatti che confermassero, per quanto era possibile, l'esistenza di atomi di determinate dimensioni fisiche. Nel corso di questa ricerca, in base alla teoria atomica, scoprii che ogni microscopica particella sospesa in un liquide; doveva presentare un particolare movimento accessibile all'osservazione, senza sapere che le osservazioni sperimentali relative ai moti browniani erano appunto già note da lungo tempo». Torneremo su queste ricerche, eccezionali per le doti sia matematicoformali, sia fisico-intuitive rivelate in questo settore dal giovane Einstein: si pensi che i risultati di Boltzmann e di Gibbs vengono considerati oggi tra i più difficili e importanti della fisica moderna.

Einstein ci ha ricordato che uno dei motori dello scienziato è la fede in una unità razionale del mondo, unità che almeno in parte è percepibile dalla nostra mente. Tra parentesi, la scienza cosiddetta «moderna» è nata (o rinata) in Europa forse perché qui il terreno vi fu in parte preparato (oltre che dalle importanti attività artigianali) dagli studi medioevali della patristica e della scolastica, ispirati alla concezione di un Dio unico. Infatti, per giudicare una teoria, Einstein spesso si domandava se - qualora fosse stato Dio - avrebbe creato l'universo in quel modo. E' ben noto, ad esempio, come (nel rifiutare la visione probabilistico-quantistica dei fenomeni microfisici fornita dalla meccanica quantistica) dichiarasse più volte di non credere che «Dio giocasse ai dadi»; e così via.

Abbiamo già visto quanto rilievo abbia dato Einstein al processo e «intuitivo» rispetto a quello «induttivo». Ritorniamo su questo aspetto dell'insegnamento einsteiniano riportando alcune delle parole da lui pronunciate nel 1918 in onore di Planck (8): «Il compito supremo dello scienziato è di pervenire a quelle leggi elementari universali partendo dalle quali il cosmo può essere costruito con la pura deduzione. Non esiste alcun sentiero logico che conduce a queste leggi: soltanto l'intuizione, appoggiata ad una sensata comprensione della realtà, può condurre ad esse... L'anelito a contemplare l'armonia cosmica è la fonte della pazienza inesauribile e della perseveranza con le quali Planck si è dedicato ai problemi più generali della fisica... Lo sforzo quotidiano scaturisce non già da un'intenzione deliberata o da un programma, bensì direttamente dal cuore». Completiamo questo quadro, menzionando - da un lato - che Einstein sottolineò nettamente anche il ruolo del «background» filosofico (ha scritto (7): «Il tipo di ragionamento critico necessario per la scoperta del carattere relativo della simultaneità mi fu reso enormemente più facile dalla lettura degli scritti filosofici di David Hume e di Ernst Mach») e - dall'altro - che Einstein chiarì come il comportamento dello studioso non sia però riconducibile ad alcun modello epistemologico prefissato (7): «E' inevitabile che lo scienziato appaia all'epistemologo sistematico come una specie di opportunista senza scrupoli: che gli appaia un realista, perché cerca di descrivere il mondo indipendentemente dagli atti di percezione; come un idealista, poiché considera i concetti e le teorie come libere invenzioni dallo spirito umano (non deducibili logicamente dal lato empirico); come un positivista, poiché ritiene che i suoi concetti e le sue teorie sono giustificati soltanto in quanto forniscono una rappresentazione logica delle relazioni fra esperienze sensoriali. Può addirittura sembrargli un platonico o un pitagorico in quanto considera il criterio della semplicità logica come strumento indispensabile ed efficace per la sua ricerca».

Non dobbiamo credere che la metodologia della scoperta scientifica conceda troppo all'irrazionale. L'intuizione serve, ad es., come mezzo per arrivare a concetti e , principi fondamentali - assunti prima come ipotesi provvisorie e poi come postulati formali da cui dedurre una spiegazione teorica del reale, da confrontare infine con l'esperienza — Questo procedere riconosce l'importanza del «pensiero puro»; la scelta della teoria e dei suoi principi fondamentali si ispira ai criteri dell'eleganza sostanziale e formale, della bellezza concettuale e matematica, e soprattutto dalla semplicità logica (che realizza anche l'aspetto machiano della scienza quale intelligente «Economia del Pensiero»). Dice Einstein (7): «Una teoria è tanto più convincente quanto più semplici sono le sue premesse, quanto più varie sono le cose che essa collega fra di loro, quanto più esteso è il suo campo di applicazione».

Se qualcuno si meravigliasse di sentire parlare di bellezza nel campo della fisica e della matematica, vorrei suggerire questo pensiero: il musicista creatore, chino sui suoi fogli di note musicali durante il lavoro di composizione, apparirebbe come un arido manipolatore di simboli grafici, se noi non possedessimo le orecchie. Purtroppo l'«orecchio» per sentire le scienze è in media molto meno sviluppato, e per nulla educato.

Ci si muove tra due poli. Il polo extra-logico domina nella fase intermedia della «ricerca»: «La più bella esperienza che possiamo avere è il misterioso: si tratta dell'emozione fondamentale, la culla della vera arte e della vera scienza», dice Einstein (8), e Hoffman, professore di matematica, aggiunge a proposito del la Relatività Generale: «Allorché constatiamo quanto apparentemente vacillanti fossero i fondamenti sui quali Einstein edificò la sua teoria, possiamo soltanto meravigliarci dell'intuito che lo guidò verso il suo capolavoro. Una simile intuizione è l'essenza del genio. Non erano forse vacillanti anche i fondamenti della teoria di Newton?... E Maxwell non costruì su un pazzesco modello meccanico, giudicato da lui stesso incredibile? Mediante una sorta di divinazione, il genio sa sin dall'inizio, in modo nebuloso, qual è la meta che deve sforzarsi di raggiungere. Durante il tragitto penoso attraverso una regione sconosciuta, esso puntella la propria fiducia con argomentazioni di plausibilità che servano uno scopo non tanto logico quanto freudiano. Queste argomentazioni non devono essere necessariamente probanti finché sono utili all'impulso irrazionale, chiaroveggente, subconscio che è in realtà dominante. Invero non dovremmo aspettarci che siano probanti pello sterile senso logico, poiché l'uomo che crea una rivoluzione scientifica deve costruire basandosi su quelle stesse idee che sta per sostituire».
Ma l'altro polo, quello logico, brilla nei risultati: Eddington, appena conosciuta la teoria, scrisse nel 1916: «Sia che la Relatività Generale risulti in definitiva esatta o meno, essa esige attenzione come uno dei più meravigliosi esempi della potenza del ragionamento matematico» (8). Ed Einstein stesso ha lasciato scritto nelle sue Note Autobiografiche (7): «La teoria della Relatività Generale procede dal seguente principio: le leggi naturali devono essere espresse da equazioni che siano "invarianti in forma" dispetto al gruppo delle "trasformazioni continue" di coordinate... Il significato altamente euristico dei principi della relatività generale sta nel fatto che essi ci portano a cercare quelle equazioni che (nella loro formulazione "generale covariante") siano le più semplici possibili». E ancora, ricordandoci questa volta Galileo: «La natura è la realizzazione dei concetti matematici più semplici. Sono convinto che noi possiamo scoprire per mezzo di costruzioni puramente matematiche i concetti e le leggi che li connettono, leggi che danno la chiave per comprendere i fenomeni naturali».
Einstein ci ha ricordato pure che «il pensiero è necessario per poter capire il dato empirico, e i concetti e le categorie sono necessari al pensiero» (7), ma che esse «categorie» sono il prodotto dell'evoluzione biologica del nostro sistema nervoso centrale, e quindi variano al variare della nostra esperienza evolutiva. Ciò vale anche par categorie come quelle di «spazio» e di «tempo»; il credere a «giudizi sintetici a priori» (cosa più che giustificata all'epoca di Kant) sarebbe oggi una ingenuità.
Infine, non dobbiamo affatto ritenere che, anche senza Einstein, la fisica prima o poi si sarebbe sviluppata allo stesso modo. Come dice Piero Caldirola, così come non ci sono due pittori che rappresentino una Madonna allo stesso modo, analogamente fisici diversi da Einstein avrebbero costruito la teoria della gravitazione, ad esempio, in modo sicuramente molto diverso. Ciò non vale solo nel caso di Einstein; per rifarci al passato, se Mach avesse sviluppato formalmente le sue idee, avrebbe forse trovato la velocità caratteristica della teoria della relatività come «velocità di espansione del cosmo» (piuttosto che come velocità della luce): «Forse, se Mach avesse avuto migliore padronanza della fisica-matematica - ha scritto B. Bertotti (9) - l'evoluzione della fisica avrebbe avuto un corso assai diverso».

Abbiamo già accennato agli incredibili risultati giovanili di Einstein nel settore della meccanica statistica. Egli fece uso dell'abilità e dei risultati così acquisiti anche in seguito, per esempio nell'introdurre i «quanti di luce» onde interpretare l'effetto fotoelettrico. Il lavoro più specifico di questa serie è però un altro. Al tempo di Einstein moltissimi fisici non creàvano ancora alla reale esistenza di atomi e molecole: questo atteggiamento inasprì ad esempio le critiche fatte da molti al grande Boltzmann, il quale - depresso in parte anche a causa di tali critiche - giunse a suicidarsi (a Duino) nel 1906. Un anno prima, però, Einstein aveva già pubblicato un articolo che - divenuro noto entro pochi anni - avrebbe dimostrato senza più dubbi l'esistenza delle molecole. Infatti, desiderando individuare un fenomeno che potesse sperimentalmente e chiaramente mettere in mostra la loro esistenza e nel contempo fornire indicazioni sulle loro dimensioni, Einstein scoprì che - a causa dell'agitazione termica molecolare - un corpicciolo sufficientemente minuscolo (ma pur sempre visibile al microscopio) avrebbe dovuto apparire soggetto ad un caratteristico moto «a zig-zag». Abbiamo già detto che Einstein non sapeva che tale fenomeno - il «moto browniano» - era già stato sperimentalmente osservato. Einstein ne fornì la derivazione teorica, e ciò permise tra l'altro di valutare le dimensioni delle inolecole. Con questo lavoro diede il crisma definitivo all'effettiva esistenza di atomi e molecole.
Nel 1905 Einstein pubblicò anche due articoli che fornivano la spiegazione dell'effetto fotoelettrico (e che gli procureranno il Premio Nobel). Con tali articoli non solo riprese l'ipotesi quantistica di Planck (formulata nel 1900, ma rimasta senza alcun seguito in quei cinque anni, perché nessuno - nemmeno Planck - aveva avuto il coraggio di accettarla davvero traendone le conseguenze), ma dimostrò che la luce in tale fenomeno si comporta come se fosse costituita da fotoni, o «quanti» di luce. Einstein mise in moto, così, l'edificazione della meccanica quantistica.

Il contributo di Einstein all'edificazione della meccanica quantistica non si limita all'introduzione dei fotoni. Ad esempio, nel 1924, L. de Broglie propose nella sua tesi di dottorato che sia la luce, sia le usuali particelle materiali fossero costituite da corpuscoli accompagnati e guidati da onde. L'ipotesi pareva folle, anche perché nel caso delle particelle elementari queste onde-guida (Pur possedendo «velocità di gruppo» inferiore alla velocità $c$ della luce nel vuoto) avevano una «velocità di fase» maggiore di $c$. Einstein comprese l'importanza della tesi di de Broglie, e ne parlò favorevolmente - con la sua solita spregiudicatezza - in un suo articolo. In seguito a questa autorevole citazione, Schrödinger ebbe il coraggio di prendere sul serio la relazione di de Broglie e pervenne a scrivere la sua notevolissima equazione («di Schrddinger»), base di tutta la moccanica quantistica propriamente detta, equazione per la quale ottenne il Premio Nobel.

Non solo: l'articolo in cui Einstein menzionò de Broglie (anche de Broglie vinse poi il Premio Nobel) apparteneva ad una serie di articoli in cui (subito comprendendo l'interesse di uno scritto di Bose, fisico indiano) introduceva la «statistica quantistica» valida per le particelle a spin intero («statistica di Bose-Einstein»).

Perfino Heisenberg (altro futuro Premio Nobel), quando ebbe costruito la sua «meccanica delle matrici» - che rivelava la grande stranezza di essere non commutativa - trovò parte del coraggio che dovette avere per non buttarla nel fuoco proprio al pensiero che Einstein aveva visto giusto nell'introdurre non una nuova dinamica, ma addirittura una nuova cinematica (8).

Einstein direttamente o indirettamente contribuì dunque in modo essenziale alla costruzione della meccanica quantistica, anche se poi la considerò sempre solo come un prodotto intermedio e incompleto (allontanandosi, forse a ragione, dalla stragrande maggioranza degli altri fisici).

Nel 1916 Einstein, nel derivare in una nuova semplicissima maniera la formula di Planck «del corpo nero», introdusse in base a considerazioni elegantissime e molto generali il processo di «emissione stimolata», che costituisce il principio fondamentale del laser. Einstein, quindi, può essere ritenuto il padre anche della «luce coerente» (o luce laser), che sta trovando ora applicazioni, tanti decenni dopo.

Uno degli eterni dilemmi della scienza sta nella scelta tra una descrizione «continua» e una descrizione «discreta» della natura fisica. Einstein, pur avendo dato tanti contributi determinanti alla nascita e costruzione della meccanica quantistica (che si rifà al «discreto»), optò decisamente per il «continuo», avvicinandosi con le sue teorie relativistiche (soprattutto la relatività generale) più ad Anassagora e agli Stoici che non agli Atomisti ed Epicurei.

Cominciamo con la Relatività Speciale, pure pubblicata nell'anno favoloso 1905, all'età di ventisei anni. Questa teoria mostra malto bene il potere del ragionamento ispirato a principi semplici e generali. Se si postula che: 1) ogni osservatore inerziale veda il mondo dei fenomeni meccanici ed elettromagnetici retto dalle medesime leggi; 2) tempo e spazio siano omogenei e lo spazio sia isòtropo, segue che deve esistere una velocità $w$ invariante (cioè che appare avere il medesimo valore a tutti gli osservatori inerziali).

Dire che «il tempo à omogeneo» vuol dire postulare la covarianz (=invarianza in forma) delle leggi fisiche rispetto alle «traslazioni temporali»: cioè ammettere che un esperimento eseguito oggi obbedisca alle medesime leggi cui obbedisce un esperimento identico: eseguito, ieri o domani; e così via. E' interessante che dal postulato n. 2) segue che esistono - come confermato dall'esperienza - delle quantità (rispettivamente Energia, Quantità di Moto o «Momentum», e Momento della Quantità di Moto o «Angular momentum») che si conservano, in un sistema isolato, durante l'avvenire delle trasmutazioni fisiche.

Se la velocità invariante $w$ fosse infinita, ne seguirebbero la Relatività Galileana (10) e la fisica di Galilei-Newton. Se, invece, seguendo Einstein e le esperienze elettromagnetiche, identifichiamo $w$ con la velocità $c$ della luce, allora ne segue la Relatività Speciale (11) einsteiniana. Uno dei principali risultati della Relatività Speciale è il seguente. Prima di Einstein i fisici avevano cominciato ad accorgersi che le misure della separazione spaziale e della separazione temporale tra due eventi erano probabilmente non «assolute», ma relative all'osservatore. Dalla Relatività Speciale (RS) si deduce subito, però, che è assoluta la distanza quadridimensionale tra due eventi. La RS dunque ci insegna a costruire quantità «assolute» a partire da quantità «relative»: essa avrebbe ben potuto essere chiamata «Teoria della Assalutività» (e chissà quanto diverse sarebbero state le reazioni di tanti letterati o filosofi o artisti solo orecchianti!). Nel 1908, a Colonia, Minkowski poté dichiarare di conseguenza: «D'ora in poi lo spazio di per sé e il tempo di per sé sono destinati ad affondare completamente, nell'ombri, e soltanto una sorta di unione di entrambi può conservare un'esistenzà indipendente». (12) Einstein stesso, riferendosi alla rappresentazione geometrica dello spazio-tempo data da Minkowski, arrivò persino a dire, il 21-3-1955: «... Noi fisici siamo convinti che la distinzione tra passato, presente e futuro sia soltanto un'illusione, anche se ostinata». (8)

Tra il 1905 e il 1907 Einstein completò anche i suoi lavori intorno alla relazione $E = mc^2$, che stabilisce la proporzionalità tra massa (totale) relativistica ed energia (totale) relativistica, rendendo superflua in fisica una delle due grandezze (ad es. la massa relativistica). Questa relazione viene di solito presentata, in modo molto scorretto, come riferita a una presunta «equivalenza» di massa ed energia. Essa, al contrario, regola la trasmutabilità, per esempio, di particelle materiali in particelle di luce (cioè di materia in radiazione, e viceversa). L'importanza pratica di questa relazione è legata (sempre per esempio) al fatto che essa regola i meccanismi della fusione termonucleare che danno vita alle stelle, e fanno in particolare emettere luce e calore al nostro Sole.

Passiamo alla Relatività Generale (13 (RG), teoria completata e pubblicata da Einstein nel 1916, dopo quasi dieci anni di sforzi, all'età di trentasette anni. La fisica classica dava per scontato che la massa inerziale *m* (che entra nella legge fondamentale della meccanica classica: $F = ma$) e la massa gravitazionale *m'* (che entra nella legge newtoniama della gravitazione universale: $F = Gm'M'/r^2$) siano proporzionali, ovvero - in opportune unità di misura - identiche. Ciò è molto ben verificato dall'esperienza: tale identità fa sì, ad esempio, che tutti i corpi sulla Terra cadano con la medesima accelerazione di gravità (*g = 9,8m/s*) indipendentemente dalla loro massa. Partendo da queste semplici considerazioni, Einstein comprese che la gravitazione può interpretarsi come dovuta non tanto ad un «campo gravitazionale» sovrapposto ad uno spazio-tempo piatto e infinito, quanto ad una deformazione geometrica dello spazio-tempo stesso, che viene incurvato dalla presenza di masse, (In tal modo, l'intero cosmo poteva divenire curvo, e magari chiuso su se stesso e quindi finito). La Tersa, ad es., si muoverà intorno al Sole semplicemente perché descriverà una «retta» (o meglio una geodetica) in uno spaziotempo leggermente incurvato dalla massa del Sole. Si noti come sia essenziale qui considerare l'incurvamento del tempo, e non solo dello spazio. Così Einstein spiegò - come noto - l'avanzamento del perielio di Mercurio, la deflessione dei raggi di luce da parte di masse gravitanti, ecc. Questo ultimo risultato si può attenere, qualitativamente, in modo immediato sulla base di uno dei famosi «esperimenti concettuali» di Einstein. Il successo nel geometrizzare la gravità spingerà Einstein a cercare una «teoria unificata», che geometrizzasse anche il campo elettromagnetico. I tempi allora non erano maturi. Ma il sogno di Einstein (incompreso da gran parte dei fisici di questo secolo) non è forse lontano, ora, dalla realizzazione.

E non dimentichiamo che le equazioni einsteiniane della RG ammettono anche particolari soluzioni di tipo «*buco nero*», che tanto sono oggidì in voga. Anche se i «black-holes» (14) non esistessero, l'analisi recente fattane da S. Hawking e Coleghi ha attenuto e sta ottenendo risultati che sono con ogni probabilità tra i più importanti di tutta la storia della fisica per la loro universalità e la loro potenziale ricchezza di significato. Non bisognerebbe neppure dimenticare le «onde gravitazionali», la cui esistenza è prevista dalla RG.

Abbiamo già accennato al fatto che la RG permette di costruire «modelli» per l'intero nostro cosmo. Einstein partì dalle proprie equazioni della RG, ispirandosi ai seguenti principi: 1) che il cosmo sia spazialmente finito; 2) che da ogni punto di osservazione esso appaia come sostanzialmente lo stesso («Principio Copernicano»). Elaborò così il proprio modello di universo, modificandolo quando Hubble scoprì la legge di fuga delle galassie, legge che ci parla dell'espansione del nostro cosmo.
I modelli dinamici di Einstein, e più in generale quelli detti di Friedmann, danno ottimi risultati nella descrizione del nostro cosmo. In conclusione, con Einstein lo stesso problema cosmologico è divenuto, per la prima volta nella storia umana, una questione all'altezza delle, capacità scientifiche dell'uomo.

Adattamento on line del presente scritto a cura di: *Felice Palmeri (Centro D.I.E.A.)*